\documentclass[12pt]{article}
\pdfoutput=1
\usepackage{amsmath}
\usepackage{graphicx,psfrag,epsf}
\usepackage{enumerate}
\usepackage{natbib}
\usepackage{setspace}
\usepackage{amssymb}
\usepackage{algorithm2e}

\def\spacingset#1{\renewcommand{\baselinestretch}%
{#1}\small\normalsize} \spacingset{1}

\begin{document}

\title{Bayesian analysis of accumulated damage models in lumber reliability}

\author{Chun-Hao Yang$^1$, James V Zidek$^2$, Samuel W K Wong$^{1,*}$\\
\small
 $^1$ Department of Statistics, University of Florida, Gainesville, FL \\ \small
 $^2$ Department of Statistics, University of British Columbia, Vancouver, BC, Canada \\ \small
 $^*$ Author for correspondence: swkwong@stat.ufl.edu
}

\date{\today}

\maketitle

\spacingset{1} 

\begin{abstract}

Wood products that are subjected to sustained stress over a period of long duration may weaken, and this effect must be considered in models for the long-term reliability of lumber.  The damage accumulation approach has been widely used for this purpose to set engineering standards.  In this article, we revisit an accumulated damage model and propose a Bayesian framework for analysis. For parameter estimation and uncertainty quantification, we adopt approximation Bayesian computation (ABC) techniques to handle the complexities of the model. We demonstrate the effectiveness of our approach using both simulated and real data, and apply our fitted model to analyze long-term lumber reliability under a stochastic live loading scenario.

\end{abstract}


\newpage

\section{Introduction}\label{sect:introduction}

The long-term reliability of lumber is an important consideration in the construction of wood-based structures.  That led \citet{foschi1979discussion} to advance the development of a system for setting lumber standards with an explicit role for probability models and return periods, and other key concepts in the theory of reliability.  There are challenges associated with the application of structural reliability to wood products, as lumber has considerable inherent variability and is susceptible to the  ``duration of load (DOL)'' effect.  The DOL effect was first studied empirically by \citet{wood1951relation}.  Briefly, when a piece of lumber is subject to a sustained stress over a period of long duration, the stress may cause it to first deform (known as \emph{creep}) and then eventually to fail (known as \emph{creep rupture}).  Thus for structural engineering applications, the DOL effect has to be taken into consideration when calculating safety factors. \citet{foschi1989reliability} conduct an in-depth study of this nature and present the reliability assessment results for structural usage of lumber, where the stress on an individual piece may be a combination of random (e.g.~from snow or owner occupancy) and constant (e.g.~from the dead weight of structure) loadings over time.

The time to failure of lumber products with a long intended life span (e.g.~30 or more years) cannot be measured for practical reasons, so instead various accelerated testing methods have been developed to study the DOL effect \citep{barrett1978duration2}.  These tests are described in terms of the load applied over time $\tau(t)$, $t\ge0$.  Two such commonly used loading patterns are the \emph{ramp load} and \emph{constant load}.  For a ramp load test the load is applied at a linearly increasing rate $\tau(t) = kt$ until the piece breaks, where $k$ is the selected loading rate in $psi$ (pounds per square inch) per unit of time.  One particular ramp loading rate $k_s$ is set for calibration purposes, corresponding to the way the short-term strength of the piece $\tau_s$ is defined: letting random variable $T_s$ denote the breaking time of the piece when ramp load rate $k_s$ is used, then $\tau_s \equiv k_s T_s$.  In contrast, the constant load test is based on applying a constant load $\tau_c$ over time:  the procedure begins with an initial ramp loading phase that increases the load to the preset level $\tau_c$, after which the test continues under that constant load.  The test ends when either the piece breaks, or the piece has survived a specified time period without breaking.  For practical purposes, the time period after which a constant load test is truncated is usually a few months to a few years.

To model the DOL effect and project the results obtained from accelerated tests to longer time periods, the damage accumulation approach has received substantial attention \citep[e.g.,][]{foschiyao1986dol,gerhards1987cumulative,rosowsky1992reliability,hoffmeyer2007duration,svensson2009duration,li2016reliability}.  In this context, $\alpha(t)$ denotes the damage state of the piece as a function of time, such that $\alpha=0$ indicates no damage and $\alpha=1$ indicates failure.  While $\alpha(t)$ is generally a latent function and not directly measurable -- as we only observe $\alpha(0) = 0$ and $\alpha(T) = 1$ where $T$ is the piece-specific random failure time -- the construction of theoretical models for $\alpha(t)$ has nonetheless served as a useful device for fitting experimental data.  A key feature of these models is that they provide a corresponding theoretical damage accumulation curve $\alpha(t)$ for any input loading profile $\tau(t)$ desired.

Accumulated damage models (ADMs) express the rate of damage accumulation in terms of a differential equation that involves $\tau(t)$ and $\tau_s$.  Various functional forms of ADMs have been proposed.  For example, the `US model' was introduced by \citet{gerhards1979timeC} and slightly modified by \citet{Vincent2011}, which specifies
\begin{equation*} \label{US_model}
\frac{d}{dt}\alpha(t) \mu  =\exp\left(-A+B\frac{\tau(t)}{\tau_{s}}\right)
\end{equation*}
where $A$ and $B$ are random effects for each specific piece of lumber, and $\mu$ is a constant with units `time' to ensure dimensional consistency.  The `Canadian model' was introduced by \citet{foschiyao1986dol}, and we consider the modified version with a reparametrization to ensure dimensional consistency,
\begin{eqnarray} \label{Can_model}
 	\frac{d}{dt}\alpha(t) \mu &=& [(a \tau_s )(\tau(t)/\tau_s - \sigma_0)_+]^b
  + [(c \tau_s )(\tau(t)/\tau_s - \sigma_0)_+]^n \alpha(t)
\end{eqnarray}
where $a$, $b$, $c$, $n$, $\sigma_0$ are piece-specific random effects and $(x)_+ = max(x, 0)$. Here, $\sigma_0$ serves as the stress ratio threshold in that damage starts to accumulate only when $\frac{\tau(t)}{\tau_s} > \sigma_0$.  The Canadian model was previously shown to provide a good fit to experimental data in \citet{foschi1989reliability}.  Hence our reparametrized Canadian model (Equation \ref{Can_model}) will be the main focus of this article, and to facilitate comparability with Foschi's results we set $\mu = 1$ hour and use `hours' as our time unit.

Previous work by \citet{foschiyao1986dol} and \citet{gerhards1987cumulative} have proposed  non-linear least squares and regression-based methods to estimate the parameters in these models based on constant load experimental data.  The fitted models were then applied with various stochastic loadings $\tau(t)$ to simulate real conditions such as snow loads and occupancy loads, to assess the reliability of pieces over long periods of time.  Thus the estimated ADM parameters have played a crucial role in the development of safety factors for wood-based structures.  However, due to computational complexities, appropriate statistical methods of parameter estimation have not been previously attempted for these models.  Such methods are necessary to better quantify the effect of uncertainty in parameter estimates on reliability.  The advances in modern statistical computation motivate us to revisit this problem and develop the necessary foundations on which current engineering standards can be evaluated and improved.  We adopt a Bayesian approach for inference, as it provides a coherent way to account for parameter uncertainties in the posterior distribution for future time to failure.

The remainder of the article is laid out as follows.  In Section \ref{sect:estimation} we discuss the difficulties encountered in parameter estimation for ADMs, and propose an adaptation of approximate Bayesian computation (ABC) techniques to tackle this problem.  In Section \ref{sect:simulation} we present results of our estimation procedure on simulated data to assess its effectiveness.  Analysis of a real dataset is provided in Section \ref{sect:analysis}.  In Section \ref{sect:prediction} we review how the ADMs are used for time--to--failure prediction under a live loading scenario, and apply our fitted model for that purpose.  We conclude the article with a brief discussion in Section \ref{sect:discussion}.

\section{Parameter Estimation for the Canadian ADM} \label{sect:estimation}

The parameter estimation problem of primary interest here is the scenario where a random sample of pieces is subject to the load profile
\begin{equation}\label{eq:loadprof}
\tau(t) = \begin{cases}
kt, \mbox{~~~~for } t \le T_0 \\
\tau_c, \mbox{~~~~for } t > T_0 \end{cases}
\end{equation}
where $\tau_c$ is the selected constant-load level, and $T_{0}={\tau_{c}}/{k}$ is the time required for the load to reach $\tau_c$ under the ramp-loading rate $k$.  For calibration purposes, the test is run with $k=k_s$ to match the ramp-loading rate used to define the short-term strength of a piece of lumber (see Introduction). The load profile in Equation (\ref{eq:loadprof}) is the general constant--load test, and includes the ramp--load test as a special case which is obtained by setting $\tau_c = +\infty$.  We first construct the likelihood function of the model parameters based on the observed data  $\boldsymbol{t}_{obs}$ for the failure times in the sample.

When we set $k = k_s$ and $\tau_c = +\infty$ for a ramp--load test, $T_s$ can be determined as a function of the piece-specific random effects. For the Canadian ADM (\ref{Can_model}), $T_s$ can only be solved numerically;  it can be shown that $T_{s}$ is determined by the solution to the equation (see Appendix)
\begin{equation}\label{eq:Tssoln}
H(T_{s})=\frac{\left(akT_{s}\right)^{b}}{\left(ckT_{s}\right)^{n(b+1)/(n+1)}}\left(\frac{\mu(n+1)}{T_{s}}\right)^{\frac{b-n}{n+1}}\int_{0}^{-\log H(T_{s})}e^{-u}u^{(b+1)/(n+1)-1}du
\end{equation}
where
\[
H(t)=\exp\left\{ -\frac{1}{\mu}\left(ckT_{s}\right)^{n}\frac{T_{s}}{n+1}\left(\frac{t}{T_{s}}-\sigma_{0}\right)^{n+1}\right\} .
\]
This provides an implicit solution of $T_s$ as a function of $a$, $b$, $c$, $n$, $\sigma_0$.

The constant--load test with the same ramp-loading rate $k=k_s$ for the initial portion ($t \le T_0$) then has a failure time $T_c$ that can be expressed in terms of $T_s$ and the piece-specific random effects,
\[
T_{c}=-\frac{1}{C_{2}}\log\left(\frac{\frac{C_{1}}{C_{2}}H^{\star}(T_{0})+C_{3}}{1+\frac{C_{1}}{C_{2}}}\right)
\]
where
\begin{eqnarray*}
C_{1} & = & \frac{1}{\mu}\left[akT_{s}\left(\frac{T_{0}}{T_{s}}-\sigma_{0}\right)\right]^{b}\\
C_{2} & = & \frac{1}{\mu}\left[ckT_{s}\left(\frac{T_{0}}{T_{s}}-\sigma_{0}\right)\right]^{n}\\
C_{3} & = & \alpha(T_{0})H^{\star}(T_{0})\\
H^{\star}(T_{0}) & = & \exp\left\{ -C_{2}T_{0}\right\} \\
\alpha(T_{0}) & = & \frac{1}{H(T_{0})}\frac{\left(akT_{s}\right)^{b}}{\left(ckT_{s}\right)^{n(b+1)/(n+1)}}\left(\frac{\mu(n+1)}{T_{s}}\right)^{\frac{b-n}{n+1}}\int_{0}^{-\log{H(T_{0}})}e^{-u}u^{(b+1)/(n+1)-1}du.
\end{eqnarray*}
Thus the complete solution for the constant-load failure time $T$ is
\begin{equation} \label{eqn:sol}
T=\begin{cases}
T_{s} & \text{if }T_{s} \leq T_{0}\\
T_{c} & \text{if }T_{s}>T_{0}
\end{cases}.
\end{equation}

To model piece-to-piece variation, it is customary to assign distributions to the piece-specific random effects $a$, $b$, $c$, $n$, $\sigma_0$, either Normal or log-Normal.  In this paper we specify these as follows:
\begin{eqnarray} \label{eqn:random-effect}
a|\mu_{a},\sigma_{a} & \sim & \text{Log-Normal}(\mu_{a},\sigma_{a}) \nonumber \\
b|\mu_{b},\sigma_{b} & \sim & \text{Log-Normal}(\mu_{b},\sigma_{b}) \nonumber \\
c|\mu_{c},\sigma_{c} & \sim & \text{Log-Normal}(\mu_{c},\sigma_{c})\\
n|\mu_{n},\sigma_{n} & \sim & \text{Log-Normal}(\mu_{n},\sigma_{n}) \nonumber \\
\eta|\mu_{\sigma_{0}},\sigma_{\sigma_{0}} & \sim & \text{Log-Normal}(\mu_{\sigma_{0}},\sigma_{\sigma_{0}}) \text{ and set } \sigma_{0} = \frac{\eta}{1+\eta}.\nonumber
\end{eqnarray}
Therefore the parameter vector of interest is $\theta = (\mu_{a}, \sigma_{a}, \mu_{b}, \sigma_{b}, \mu_{c}, \sigma_{c}, \mu_{n}, \sigma_{n}, \mu_{\sigma_{0}}, \sigma_{\sigma_{0}})$.

The likelihood contribution of one observation $T=t$ would be
\begin{eqnarray*}
f_T(t|\theta) & = & \idotsint p(t | a, b, c, n, \sigma_0) p(a, b, c, n, \sigma_0|\theta)\,da\,db\,dc\,dn\,d\sigma_0\\
              & = & \idotsint I_{\lbrace (a,b,c,n,\sigma_0): h(a,b,c,n,\sigma_0) = t \rbrace} p(a, b, c, n, \sigma_0|\theta)\,da\,db\,dc\,dn\,d\sigma_0
\end{eqnarray*}
where $h$ is the implicit solution of $T$ expressed in terms of $a$, $b$, $c$, $n$, $\sigma_0$. The set $\lbrace (a,b,c,n,\sigma_0): h(a,b,c,n,\sigma_0) = t \rbrace$ involves $h$ which has no closed form. Therefore in practice one cannot directly calculate the likelihood using this integral.

In addition, the constant-load test is truncated after a certain period of time for practical reasons, so the likelihood of the observed data $t_{obs}$ for one board would be
\begin{equation} \label{eqn:likelihood}
f(t_{obs}|\theta) = f_T(t_{obs}|\theta)I_{[t \leq t_c]} + \left(1-F_T(t_c|\theta)\right)I_{[t > t_c]}
\end{equation}
where $t_c$ is the censoring time and $f_T(\cdot | \theta)$ and $F_T(\cdot | \theta)$ are respectively the density function and the distribution function of $T$ determined by the solution (\ref{eqn:sol}).

Since $f_T(t | \theta)$ is intractable, so is the likelihood $f(t_{obs}|\theta)$. To bypass the intractability of the likelihood calculation, one brute-force approach that can achieve any required level of approximation accuracy is to generate a large number of $T$'s using (\ref{eqn:sol}) given $\theta$.   The density can then be approximated based on the simulated $T$'s, for example, using kernel density estimation. This poses no conceptual difficulty since it is simple to generate $a$, $b$, $c$, $n$, $\sigma_0$ from (\ref{eqn:random-effect}) and then with these random effects, we can solve for $T$ using (\ref{eqn:sol}) numerically. However, a parameter estimation procedure, if we wish to maximize the likelihood, will require many probability density evaluations which makes this approach impractical. Analytical gradient-based or convex optimization methods are also unavailable for this likelihood function, which rules out the possibility of direct maximization.

The estimated parameters will finally be used in the context of constructing time--to--failure distributions under simulated live loadings.  To propagate uncertainty in the parameter estimates to those distributions in a statistically coherent way, we adopt a Bayesian approach for inference on $\theta$ based on Markov Chain Monte Carlo (MCMC) simulation.  MCMC is also an appealing approach for exploring the parameter space without requiring gradients.  Nevertheless, a vanilla MCMC algorithm also requires repeated likelihood computation at each iteration. In what follows, we adopt the approximate Bayesian computation (ABC) technique as a likelihood-free version of MCMC, which only requires the ability to generate data from the model; thus, this allows us to sample $\theta$ in an efficient way. In Section \ref{sub:Introduction-to-ABC}, we briefly review ABC as based on the MCMC algorithm described in \cite{fearnhead2012constructing}. In Section \ref{sub:Modifications}, we propose suitable modifications to the algorithm to handle the censoring in our data. Some implementation details are provided in Section \ref{sub:choice}.

\subsection{Review of ABC-MCMC}  \label{sub:Introduction-to-ABC}

Let $\theta$ be the parameter vector of interest and $\boldsymbol{y}_{obs}$ be the $n$-dimensional observed data. The key step in ABC is the approximation of the posterior
\[
\pi(\theta|\boldsymbol{y}_{obs})\approx\pi_{ABC}(\theta|\boldsymbol{s}_{obs})\propto\pi(\theta)p(\boldsymbol{s}_{obs}|\theta)
\]
where $\boldsymbol{s}_{obs}=S(\boldsymbol{y}_{obs})$ for some summary statistics $S(\cdot)$. Then $p(\boldsymbol{s}_{obs}|\theta)$ is defined via a further approximation step,
\[
p(\boldsymbol{s}_{obs}|\theta)=\int\pi(\boldsymbol{y}|\theta)K_{\delta}(S(\boldsymbol{y})-s_{obs})d\boldsymbol{y}
\]
where $K_{\delta}(\cdot)$ is a density kernel with bandwidth $\delta>0$ \citep{fearnhead2012constructing}. Hence an MCMC sampling algorithm for this ABC posterior is given in Algorithm \ref{algo1}, where $g$ is a specified proposal distribution.

\begin{algorithm}
\caption{ABC-MCMC sampling algorithm of \cite{fearnhead2012constructing}}
\label{algo1}
\begin{enumerate}

\item Generate $\theta^{\prime}$ from $g(\theta|\theta_{k})$
\item Generate $\boldsymbol{y}$ from $f(\boldsymbol{y}|\theta^{\prime})$ and find $\boldsymbol{s}=S(\boldsymbol{y})$
\item Calculate
\[
\alpha(\theta^{\prime},\theta_{k})=\min\left(1,\frac{K_{\delta}(\boldsymbol{s}-\boldsymbol{s}_{obs})\pi(\theta^{\prime})g(\theta^{\prime}|\theta_{k})}{K_{\delta}(\boldsymbol{s}_{k}-\boldsymbol{s}_{obs})\pi(\theta_{k})g(\theta_{k}|\theta^{\prime})}\right)
\]
\item Accept $\theta^{\prime}$ and $\boldsymbol{s}$ with probability $\alpha(\theta^{\prime},\theta_{k})$; otherwise $\theta_{k+1}=\theta_{k}$ and $\boldsymbol{s}_{k+1} = \boldsymbol{s}_{k}$
\end{enumerate}
\end{algorithm}

\subsection{Modified ABC-MCMC for censored data}  \label{sub:Modifications}

From previous research (e.g. \cite{joyce2008approximately}; \cite{fearnhead2012constructing}; \cite{beaumont2002approximate}), the choice of summary statistics plays a crucial role in the success of an ABC algorithm. Ideally, if the summary statistics $S(\cdot)$ are sufficient for $\theta$, then the ABC posterior is identical to the true posterior. For most real applications it is impossible to find such a sufficient statistic, and so $S(\cdot)$ would be chosen to contain as much information about $\theta$ as possible while being of fairly low dimension. In our model, the number (or proportion) of censored observations is certainly an informative statistic, but on a different scale than other summary statistics such as means or quantiles. This problem could potentially be solved by designing an appropriate metric that combines statistics computed from the censored and uncensored observations. But here we can instead exploit a feature special to our context and factorize the likelihood to reduce the scope of the density kernel approximation to the uncensored observations.

Based on the likelihood (\ref{eqn:likelihood}), the joint likelihood of the $n$-dimensional observation $\boldsymbol{t}_{obs}$ of an iid sample is
\begin{eqnarray}\label{eq:onesamplik}
f(\boldsymbol{t}_{obs}|\theta) & = & f_T(\boldsymbol{t}_{obs}^{\prime}|\theta)\left[1-F_T(t_c|\theta)\right]^{n_{c}}\\ \nonumber
							  & = & \left[\prod_{i=1}^{n-n_c}f_T(t_{obs, i}^{\prime}|\theta)\right] \left[1-F_T(t_c|\theta)\right]^{n_{c}}\\ \nonumber
							  & = & \left[\prod_{i=1}^{n-n_c}\frac{f_T(t_{obs, i}^{\prime}|\theta)}{F_T(t_c|\theta)}\right] \left[F_T(t_c|\theta)\right]^{n-n_c} \left[1-F_T(t_c|\theta)\right]^{n_{c}}
\end{eqnarray}
where by exchangeability $n_{c}$ is the number of censored pieces in the observed data and $\boldsymbol{t}_{obs}^{\prime}$ is the uncensored part of the observed data. Since $\frac{f_T(t|\theta)}{F_T(t_c|\theta)}$, $0 \le t \le t_c$ is a normalized density, the density kernel approximation when applied to $\prod_{i=1}^{n-n_c}\frac{f_T(t_{obs, i}^{\prime}|\theta)}{F_T(t_c|\theta)}$ will have correct scaling for any choice of summary statistics on $\boldsymbol{t}_{obs}^{\prime}$.  This yields the posterior
\begin{eqnarray*}
\pi(\theta|\boldsymbol{t}_{obs}) & \propto & f(\boldsymbol{t}_{obs}|\theta)\pi(\theta)\\
 & = & \left[\prod_{i=1}^{n-n_c}\frac{f_T(t_{obs, i}^{\prime}|\theta)}{F_T(t_c|\theta)}\right] \left[F_T(t_c|\theta)\right]^{n-n_c} \left[1-F_T(t_c|\theta)\right]^{n_{c}} \pi(\theta)\\
 & \propto & \pi(\theta|\boldsymbol{t}_{obs}^{\prime}) \left[F_T(t_c|\theta)\right]^{n-n_c} \left[1-F_T(t_c|\theta)\right]^{n_{c}}\\
 & \approx & \pi_{ABC}(\theta|\boldsymbol{s}_{obs}^{\prime}) \left[F_T(t_c|\theta)\right]^{n-n_c} \left[1-F_T(t_c|\theta)\right]^{n_{c}}
\end{eqnarray*}
where $\boldsymbol{s}_{obs}^{\prime}=S(\boldsymbol{t}_{obs}^{\prime})$, and $F_T(t_c|\theta)$ can be estimated consistently by $\hat{F}_T(t_c|\theta)=n^{-1}\sum_{i=1}^{n}I_{[t_{i}<t_c]}$ using the simulated $t_{i}$ from $f_T(t|\theta)$ that will already be generated as part of an ABC-MCMC algorithm.

This provides an approximation of the Metropolis-Hastings (M-H) acceptance ratio. We present Algorithm \ref{algo2}, as a modified version of Algorithm \ref{algo1}, which we use throughout this paper.

\begin{algorithm}
\caption{ABC-MCMC sampling algorithm for censored data}
\label{algo2}
\begin{enumerate}
\item Generate $\theta^{\prime}$ from $g(\theta|\theta_{k})$
\item Generate $\boldsymbol{t} = (t_1,...,t_n)$ from $f_T(t|\theta^{\prime})$ and truncate the data with the censoring level $t_c$
\item Calculate $\boldsymbol{s} = S(\boldsymbol{t}^{\prime})$ and the censored proportion $\hat{p}_{t_c}=n^{-1}\sum_{i=1}^{n}I_{[t_{i}<t_c]}$ where $\boldsymbol{t}^{\prime}$ is the uncensored part of the simulated data
\item Calculate
\[
\alpha(\theta^{\prime},\theta_{k})=\min\left(1,\frac{K_{\delta}(\boldsymbol{s} - \boldsymbol{s}_{obs}^{\prime})\pi(\theta^{\prime})g(\theta^{\prime}|\theta_{k})}{K_{\delta}(\boldsymbol{s}_{k}-\boldsymbol{s}_{obs}^{\prime})\pi(\theta_{k})g(\theta_{k}|\theta^{\prime})}\left(\frac{1-\hat{p}_{t_c}}{1-\hat{p}_{t_c,k}}\right)^{n-n_{c}}\left(\frac{\hat{p}_{t_c}}{\hat{p}_{t_c,k}}\right)^{n_{c}}\right)
\]
\item Accept $(\theta^{\prime}, \boldsymbol{s}, \hat{p}_{t_c})$ with probability $\alpha(\theta^{\prime},\theta_{k})$; otherwise $(\theta_{k+1}, \boldsymbol{s}_{k+1}, \hat{p}_{t_c, k+1}) = (\theta_{k}, \boldsymbol{s}_{k}, \hat{p}_{t_c, k})$

\end{enumerate}
\end{algorithm}

\subsection{Choice of summary statistics, kernel and bandwidth $\delta$} \label{sub:choice}

To implement Algorithm \ref{algo2}, we need to first choose the summary statistic $S(\cdot)$, the kernel function $K(\cdot)$ and the bandwidth $\delta$. For our model, there is clearly no natural sufficient statistic for $\theta$. Hence, for implementation we choose 19 equally spaced quantiles from $5\%$ to $95\%$ as the summary statistics for the uncensored part and a Normal kernel; the use of quantiles as summary statistics was previously suggested in \cite{allingham2009bayesian}. To determine the bandwidth $\delta$, we run several short simulations with different values of $\delta$ and choose the smallest one that attains a 1\% acceptance rate as suggested by \cite{fearnhead2012constructing}.

\section{Simulated Examples} \label{sect:simulation}

In this section, we set up simulation studies to illustrate the proposed Algorithm \ref{algo2}. First, we choose the following fairly diffuse prior distributions for most of the parameters in $\theta$, which will be used for both the simulated examples as well as the following real data analysis,
\begin{eqnarray*}
\mu_{a},\mu_{b},\mu_{c},\mu_{n} & \sim & N(0,20)\\
\sigma_{a}^{2},\sigma_{b}^{2},\sigma_{c}^{2},\sigma_{n}^{2},\sigma_{\sigma_{0}}^{2} & \sim & \text{Inv-Gamma}(0.01,0.01).
\end{eqnarray*}
The exception is $\mu_{\sigma_{0}}$, for which we set the more informative prior $\mu_{\sigma_{0}} \sim N(0,1)$ to correspond with the \emph{a priori} belief that on average no damage accumulates on a piece of lumber until the stress level exceeds 50\% of its short-term strength $\tau_s$ \citep[p.181]{smith2003fracture}.

The two simulation scenarios we demonstrate here are (i) fitting one constant-load dataset, and (ii) fitting two datasets with different constant-load levels simultaneously. Throughout these two simulation scenarios, the procedure is as follows: we draw a sample with size $N$ which is chosen to be close to the real data sample size. The proposal density $g(\theta^{\prime}|\theta_{k})$ for the random-walk Metropolis-Hastings is $N(\theta_{k}, \Sigma)$, where
$\Sigma = diag\{0.01, 0.01, 0.01, 0.01, 0.2, 0.01, 0.01, 0.01, 0.1, 0.01\}$. We generate $N$ observations with $\theta=(-7.50, 0.50, 3.20, 0.20, -22.00, 0.30, -1.00, 0.20, 0.15, 0.05)$.  These parameter values produce data that are somewhat similar to the real data.  We then run Algorithm \ref{algo2} to obtain 500 posterior draws of $\theta$ with 100,000 burn-in iterations and thining interval 10,000.  To explore the effect of $\delta$, we run simulations for 30 different values of $\delta$ equally spaced between 0.1 and 3 and choose the one with the acceptance rate closest to $1\%$ as our choice of $\delta$.

To evaluate the quality of the sampled $\theta$'s, we can perform an approximate log-likelihood calculation on a small subset of the samples from the MCMC run, such as our thinned list of 500 posterior draws.  To do so, we generate 100,000 failure times for each given $\theta$ and use kernel density estimation to calculate the log-likelihood for $\boldsymbol{t}_{obs}$. Recall that it is not practical to calculate the log-likelihood in this way during the MCMC for computing the M-H ratio, as to obtain an accurate estimate of the density we need to generate a large number of observations (e.g.~100,000), which is time-consuming. In contrast, for our algorithm we only need to generate the same number of observations as in the data at each MCMC iteration and hence runs very efficiently for exploring the parameter space.

\paragraph{Scenario 1: 4500/1Y} In this scenario, we generate $N=300$ observations with $\tau_c = 4500$ and the duration of the test being one year and fit this dataset with Algorithm \ref{algo2}.

\begin{figure}[!htbp]
\centering
\includegraphics[scale=1]{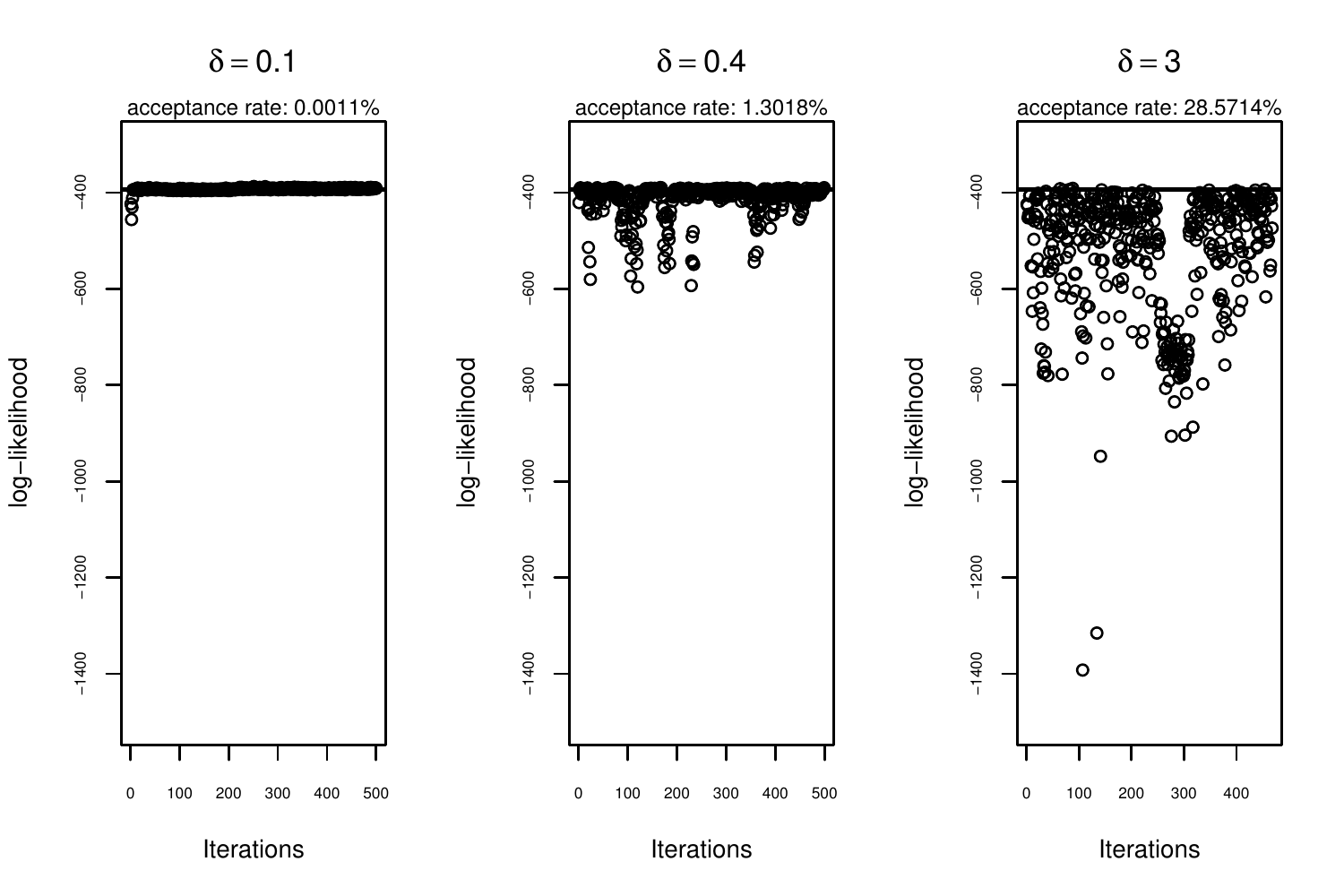}
\caption{Log-likelihoods of $t_{obs}$ for 500 ABC-MCMC samples of $\theta$ under different $\delta$; the solid line is the log-likelihood of $t_{obs}$ given the true $\theta$}
\label{fig:log-like}
\end{figure}

In Figure \ref{fig:log-like}, we have plotted these log-likelihoods for $\theta$ from three different $\delta$ values. It is clear that the choice of $\delta$ affects the quality of the simulation and goodness of the likelihood approximation.  While small values of $\delta$ theoretically provide the best approximation, the extremely low acceptance rate renders $\delta= 0.1$ to be useless in practice. For large values of $\delta$, the approximation is too crude which makes the accepted draws of $\theta$ unreliable from the likelihood perspective.  The results show that $\delta = 0.4$ indeed works well, with both a reasonable acceptance rate and good approximation to the likelihood.

\begin{figure}[!htbp]
\centering
\includegraphics[scale=1]{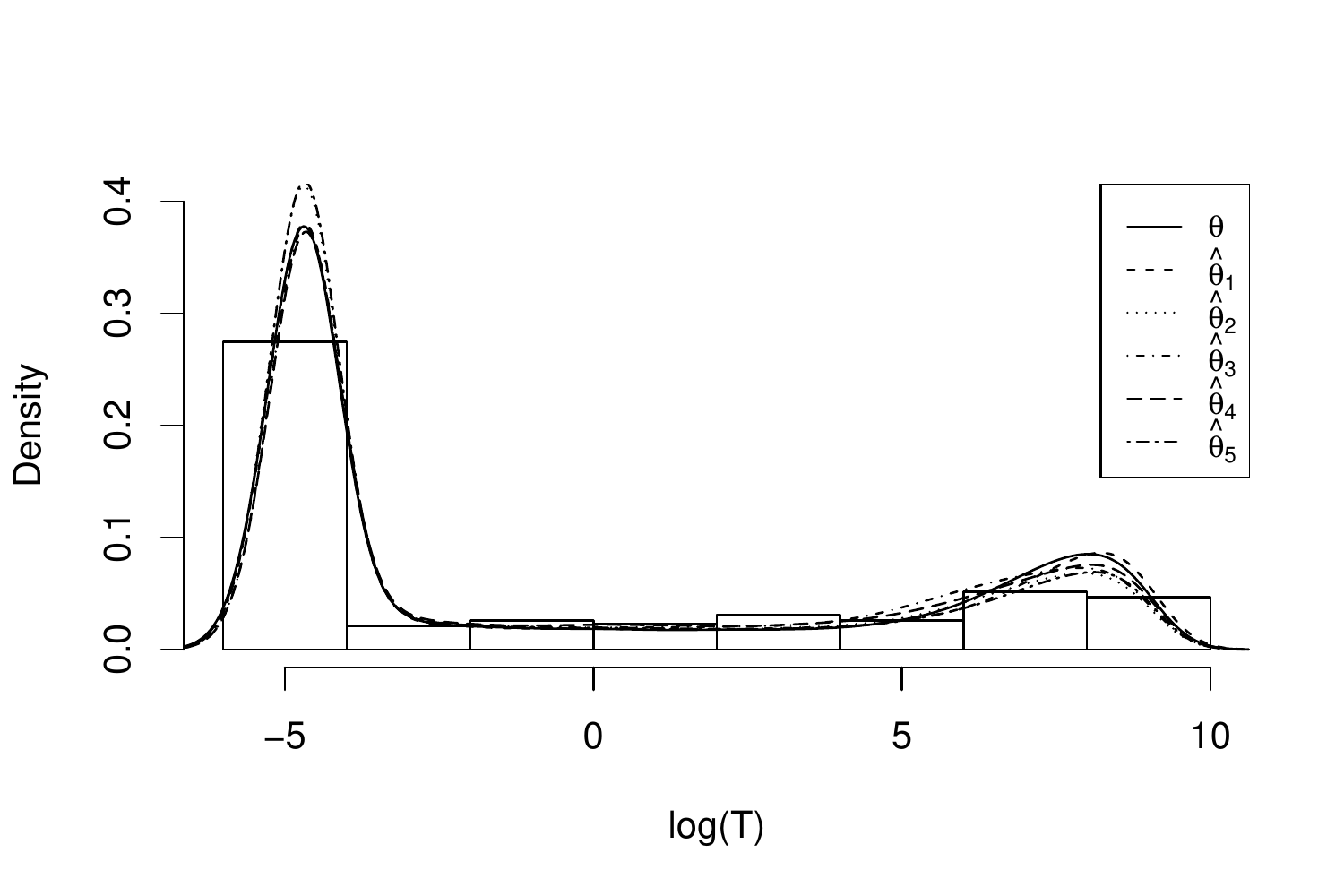}
\caption{Histogram of simulated data and estimated densities for the parameter vectors in Table \ref{table: estimation_sim1}.}
\label{fig:density_sim1}
\end{figure}

Table \ref{table: estimation_sim1} shows five sampled parameter vectors that produce the closest log-likelihood values to that of the true $\theta$. These parameter vectors are quite different yet their log-likelihoods (and log-posteriors, since the priors are mostly diffuse) are very similar. This indicates some of the model parameters are quite uncertain and the likelihood is flat over a wide range of values. Indeed, this shows that our ABC-MCMC algorithm is able to traverse the parameter space to find parameter vectors that can fit the observed data well, and so the high level of uncertainty is not a difficulty in practice for the algorithm. Figure \ref{fig:density_sim1} shows that the estimated densities $f_T(t)$ based on kernel smoothing for these parameter vectors are almost indistinguishable.

\begin{center}
\begin{table}[!htbp]
\caption{Five sampled parameter vectors, log-likelihoods, and log-posteriors for the simulated data in Scenario 1. The true value of $\theta$ is shown in the top row.}
\begin{tabular}{ c|cccccccccc|c|c}
                 & $\mu_{a}$ & $\sigma_{a}$ & $\mu_{b}$ & $\sigma_{b}$ & $\mu_{c}$ & $\sigma_{c}$ & $\mu_{n}$ & $\sigma_{n}$ & $\mu_{\sigma_0}$ & $\sigma_{\sigma_0}$ & $ll$ & log-$post$\\
\hline
$\theta$         & -7.50 & 0.50 & 3.20 & 0.20 & -22.00 & 0.30 & -1.00 & 0.20 & 0.15 & 0.05 & -393.96 & -418.27 \\
$\hat{\theta}_1$ & -8.22 & 0.45 & 3.99 & 0.10 & -42.88 & 0.12 & -1.57 & 0.34 & -1.54 & 0.71 & -389.71 & -417.46 \\
$\hat{\theta}_2$ & -7.44 & 0.43 & 3.39 & 0.62 & -30.22 & 0.55 & -1.40 & 0.32 & 0.17 & 0.20 & -389.78 & -416.29 \\
$\hat{\theta}_3$ & -7.85 & 0.44 & 3.64 & 0.38 & -38.24 & 0.18 & -1.52 & 0.50 & -0.48 & 0.11 & -389.84 & -415.97 \\
$\hat{\theta}_4$ & -8.02 & 0.43 & 3.91 & 0.58 & -13.92 & 0.15 & 0.13 & 0.42 & -0.79 & 0.24 & -390.02 & -415.26 \\
$\hat{\theta}_5$ & -7.44 & 0.44 & 3.26 & 0.40 & -30.67 & 0.62 & -1.37 & 0.27 & 0.23 & 0.28 & -390.14 & -416.51 \\
\end{tabular}
\label{table: estimation_sim1}
\end{table}
\end{center}

\paragraph{Scenario 2: 4500/1Y and 3000/4Y} In practice, multiple test samples with different constant--load levels are used to help calibrate the parameters. So we simulate such a scenario as well to see how the estimation can improve.  For this purpose, we generate a second independent dataset with $N=200$ observations, $\tau_c = 3000$, and a test duration of four years. We then fit this dataset together with the dataset in Scenario 1. In order to fit $D>1$ datasets at the same time (in this scenario, $D = 2$), we apply the ABC approximation to Equation (\ref{eq:onesamplik}) for each dataset separately.  So a corresponding minor change in the M-H acceptance ratio is needed.  Using the proposal $\theta^\prime$, we generate the multiple datasets with the different settings (i.e. $N$, $\tau_c$, and the test duration) and calculate the summary statistics and the censored proportions. To calculate the M-H acceptance ratio, we multiply together the parts involving the summary statistics and the censored proportions, i.e.
\[
\alpha(\theta^{\prime},\theta_{k})=\min\left(1,\frac{\pi(\theta^{\prime})g(\theta^{\prime}|\theta_{k})}{\pi(\theta_{k})g(\theta_{k}|\theta^{\prime})}\prod_{d=1}^{D}\frac{K_{\delta}(\boldsymbol{s}^{(d)}-\boldsymbol{s}_{obs}^{(d)})}{K_{\delta}(\boldsymbol{s}_{k}^{(d)}-\boldsymbol{s}_{obs}^{(d)})}\left(\frac{1-\hat{p}_{t_{c}}^{(d)}}{1-\hat{p}_{t_{c},k}^{(d)}}\right)^{n^{(d)}-n_{c}^{(d)}}\left(\frac{\hat{p}_{t_{c}}^{(d)}}{\hat{p}_{t_{c},k}^{(d)}}\right)^{n_{c}^{(d)}}\right)
\]
where the superscript $(\cdot)^{(d)}$ denotes the values for the $d$th simulated dataset, $d=1,...,D$.

\begin{figure}[!htbp]
\centering
\includegraphics[scale=1]{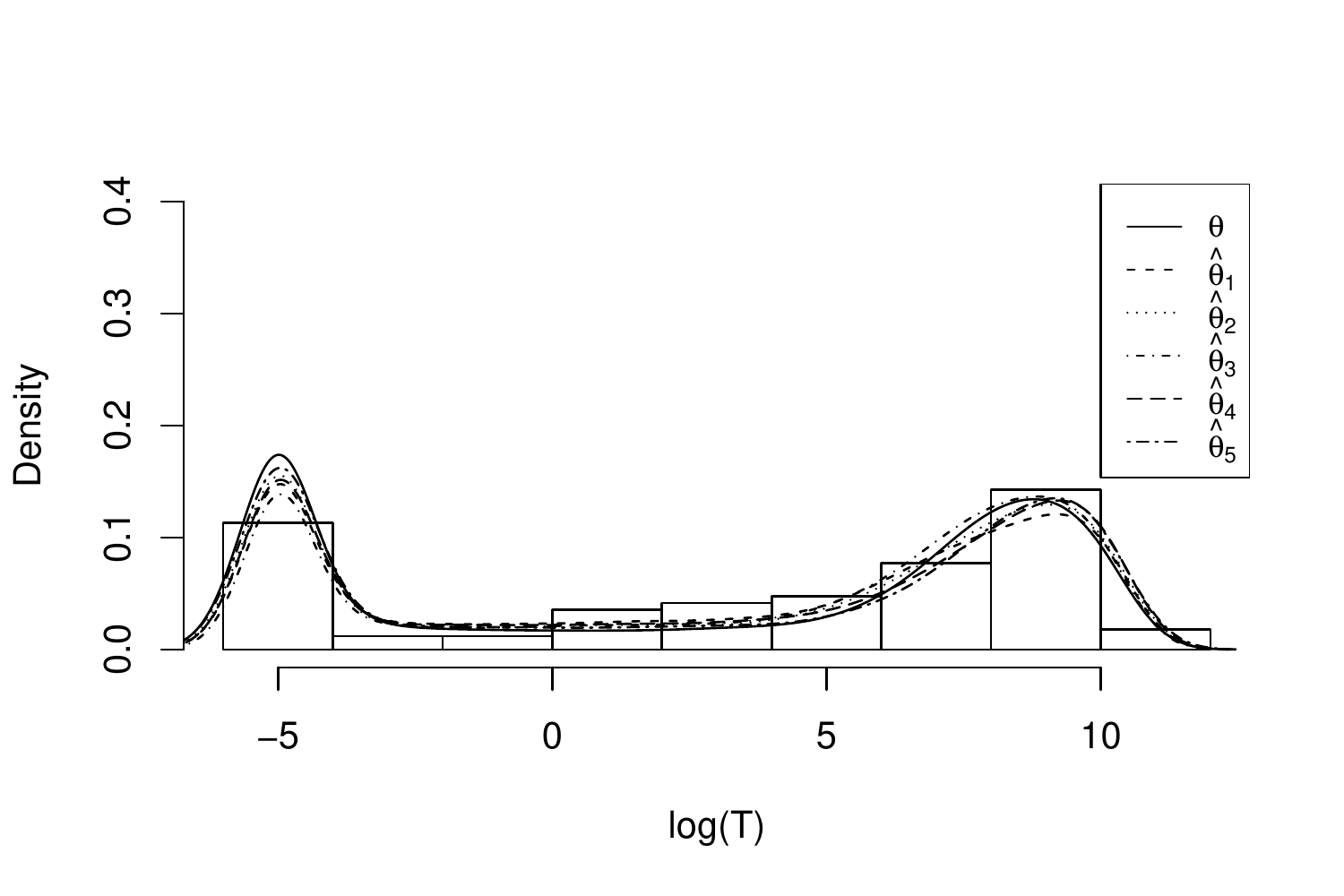}
\includegraphics[scale=1]{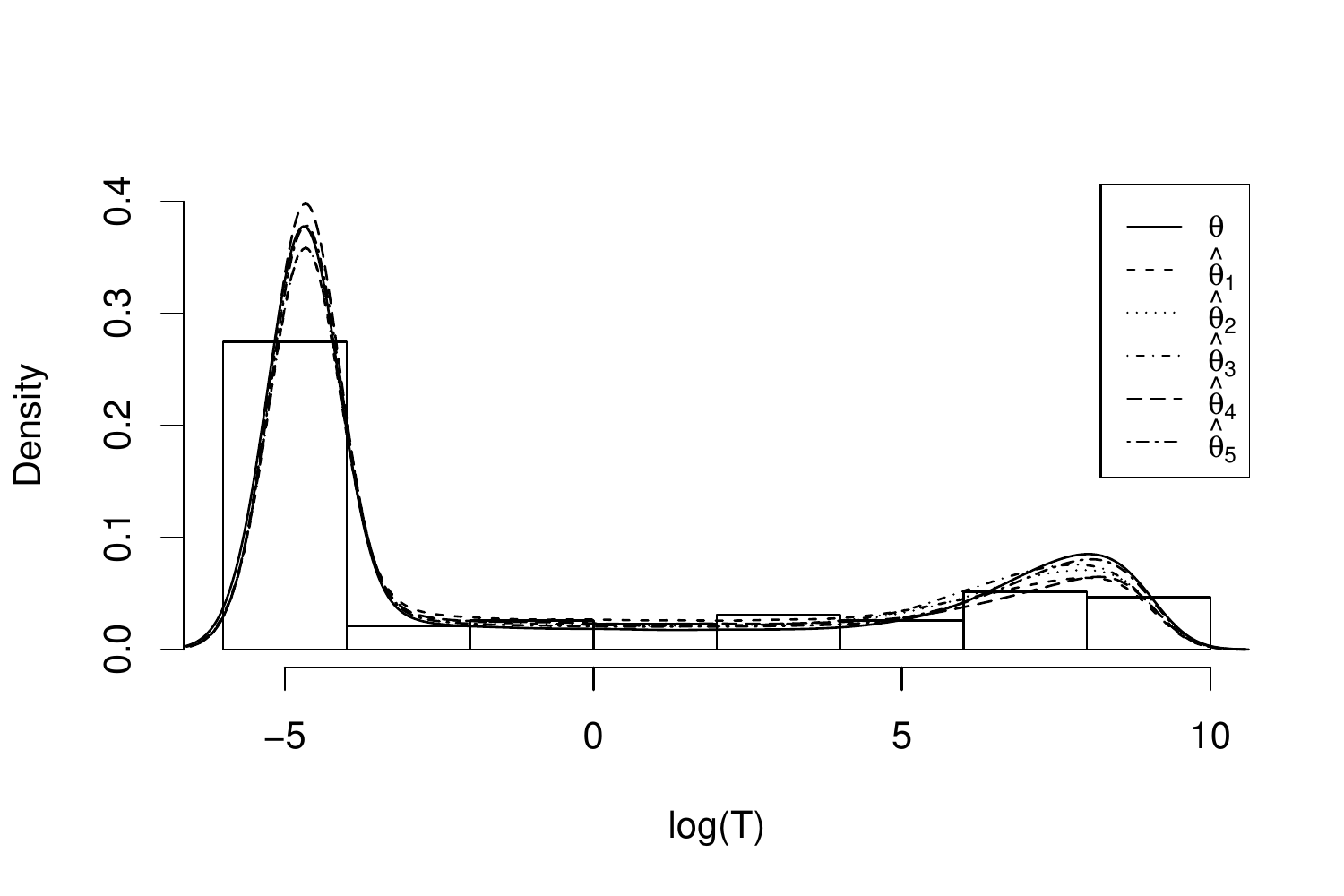}
\caption{Histograms of simulated data and estimated densities for the parameter vectors in Table \ref{table: estimation_sim2}. The top panel is the simulated 3000/4Y dataset and the bottom panel is the simulated 4500/1Y dataset.}
\label{fig:density_sim2}
\end{figure}

For this simulation, we chose $\delta = 1.1$ which gave an acceptance rate of $1.18\%$. The results are shown in Table \ref{table: estimation_sim2} and Figure \ref{fig:density_sim2}. To see how the estimation improves, we examined the standard deviations of the posterior draws of $\hat{\theta}$ and the range of their log-likelihoods. Indeed, the posterior standard deviations in Scenario 2 are smaller than those in Scenario 1;  as well, $95\%$ of the log-likelihoods in Scenario 1 lie in the range $(-494.77, -389.71)$ while $95\%$ of the log-likelihoods in Scenario 2 lie in the tighter range $(-1072.76, -1035.25)$, indicating that incorporating the additional dataset makes the estimation and our ABC algorithm more stable.

\begin{center}
\begin{table}[!htbp]
\caption{Five sampled parameter vectors, log-likelihoods, and log-posteriors for the simulated data in Scenario 2. The true value of $\theta$ is shown in the top row.}
\begin{tabular}{ c|cccccccccc|c|c}
                 & $\mu_{a}$ & $\sigma_{a}$ & $\mu_{b}$ & $\sigma_{b}$ & $\mu_{c}$ & $\sigma_{c}$ & $\mu_{n}$ & $\sigma_{n}$ & $\mu_{\sigma_0}$ & $\sigma_{\sigma_0}$ & $ll$ & log-$post$\\
\hline
$\theta$         & -7.50 & 0.50 & 3.20 & 0.20 & -22.00 & 0.30 & -1.00 & 0.20 & 0.15 & 0.05 & -1040.83 & -1065.14 \\
$\hat{\theta}_1$ & -8.09 & 0.52 & 3.48 & 0.41 & -38.56 & 0.70 & -1.21 & 0.53 & -0.98 & 0.21 & -1035.25 & -1063.41 \\
$\hat{\theta}_2$ & -7.88 & 0.48 & 3.48 & 0.25 & -17.33 & 0.24 & -0.29 & 0.46 & -0.39 & 0.20 & -1035.31 & -1060.06 \\
$\hat{\theta}_3$ & -7.88 & 0.41 & 3.47 & 0.48 & -11.08 & 0.11 & 0.42 & 0.21 & -0.43 & 0.79 & -1035.52 & -1060.64 \\
$\hat{\theta}_4$ & -8.04 & 0.45 & 3.58 & 0.26 & -24.59 & 0.27 & -0.73 & 0.41 & -0.87 & 0.65 & -1035.94 & -1062.29 \\
$\hat{\theta}_5$ & -7.57 & 0.45 & 3.28 & 0.44 & -11.55 & 0.07 & 0.26 & 0.19 & 0.08 & 0.11 & -1036.02 & -1059.60
\end{tabular}
\label{table: estimation_sim2}
\end{table}
\end{center}

\section{Data analysis} \label{sect:analysis}

The illustrative real data example comes from a duration--of--load experiment performed on visually graded 2x6 Western Hemlock, which was first analyzed in \citet{foschi1982load}.  The experimental data consists of three groups, and the standard ramp-loading rate  $k_s = 388440$psi/hour was used throughout:
\begin{enumerate}
\item A set of 300 pieces was subject to a constant-load test with $\tau_c = 4500$psi for a duration of 1 year.   In total, 56 pieces failed during the initial portion of the test, 98 failed during the 1-year constant-load period, and 146 survived to the end of the 1-year at which point the test was truncated.
\item A set of 198 pieces was subject to a constant-load test with $\tau_c = 3000$psi for a duration of 4 years.  In total, 4 pieces failed during the initial portion of the test, 42 failed during the 4-year constant-load period, and 152 survived to the end of the 4-years at which point the test was truncated.
\item A set of 139 pieces was subject to the ramp-load test, i.e.~$\tau_c = +\infty$ in Equation (\ref{eq:loadprof}).  The sample mean of short-term strength $\tau_s$ in this set was 6936psi, and sample SD 2833psi.
\end{enumerate}

\begin{figure}[!htbp]
\centering
\includegraphics[scale=1]{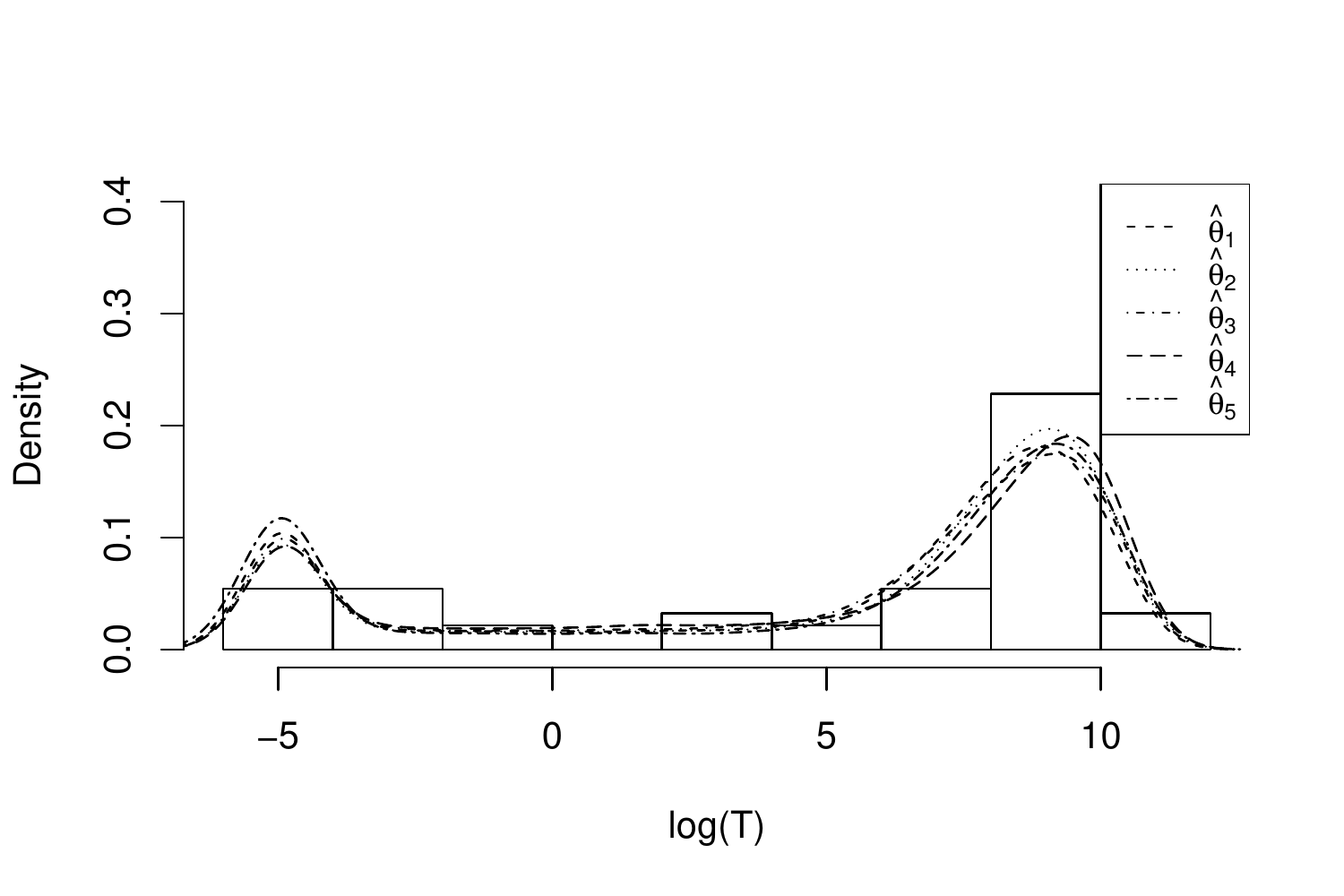}
\includegraphics[scale=1]{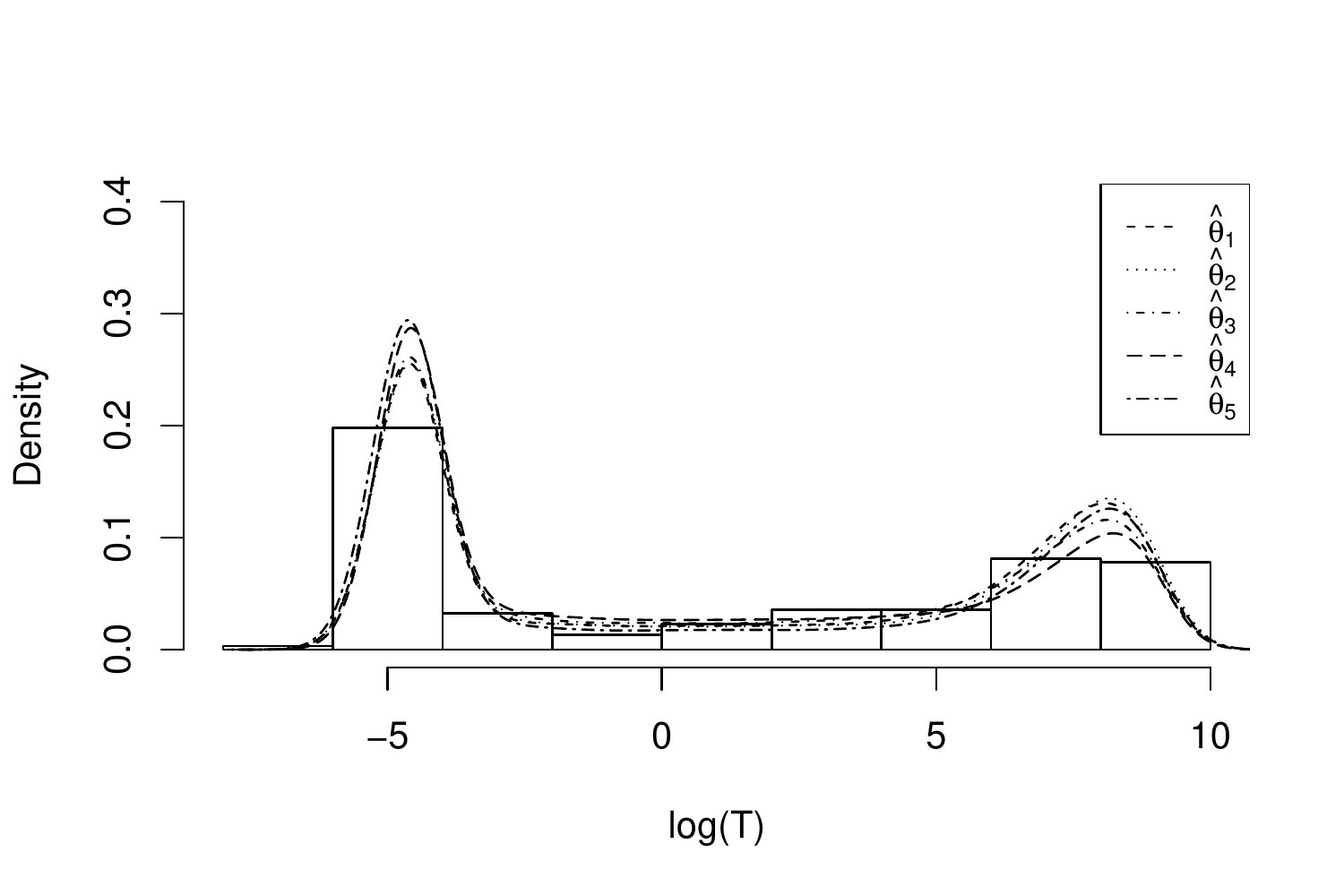}
\caption{Histograms of the real datasets and estimated densities for the parameter vectors in Table \ref{table: estimation_data}. The top panel is the 3000/4Y dataset and the bottom panel is the 4500/1Y dataset.}
\label{fig:density_data}
\end{figure}

\begin{figure}[!htbp]
\centering
\includegraphics[scale=1]{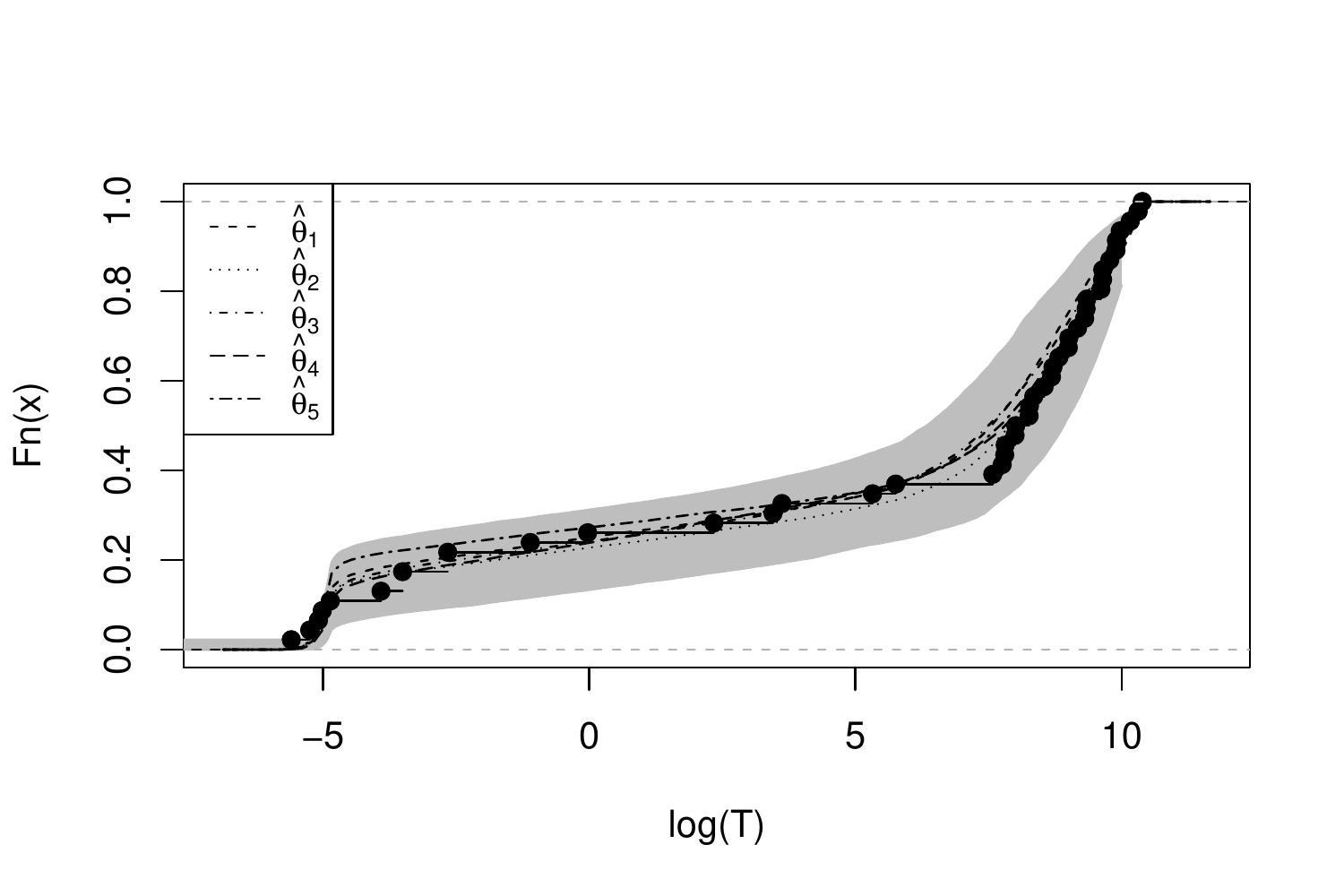}
\includegraphics[scale=1]{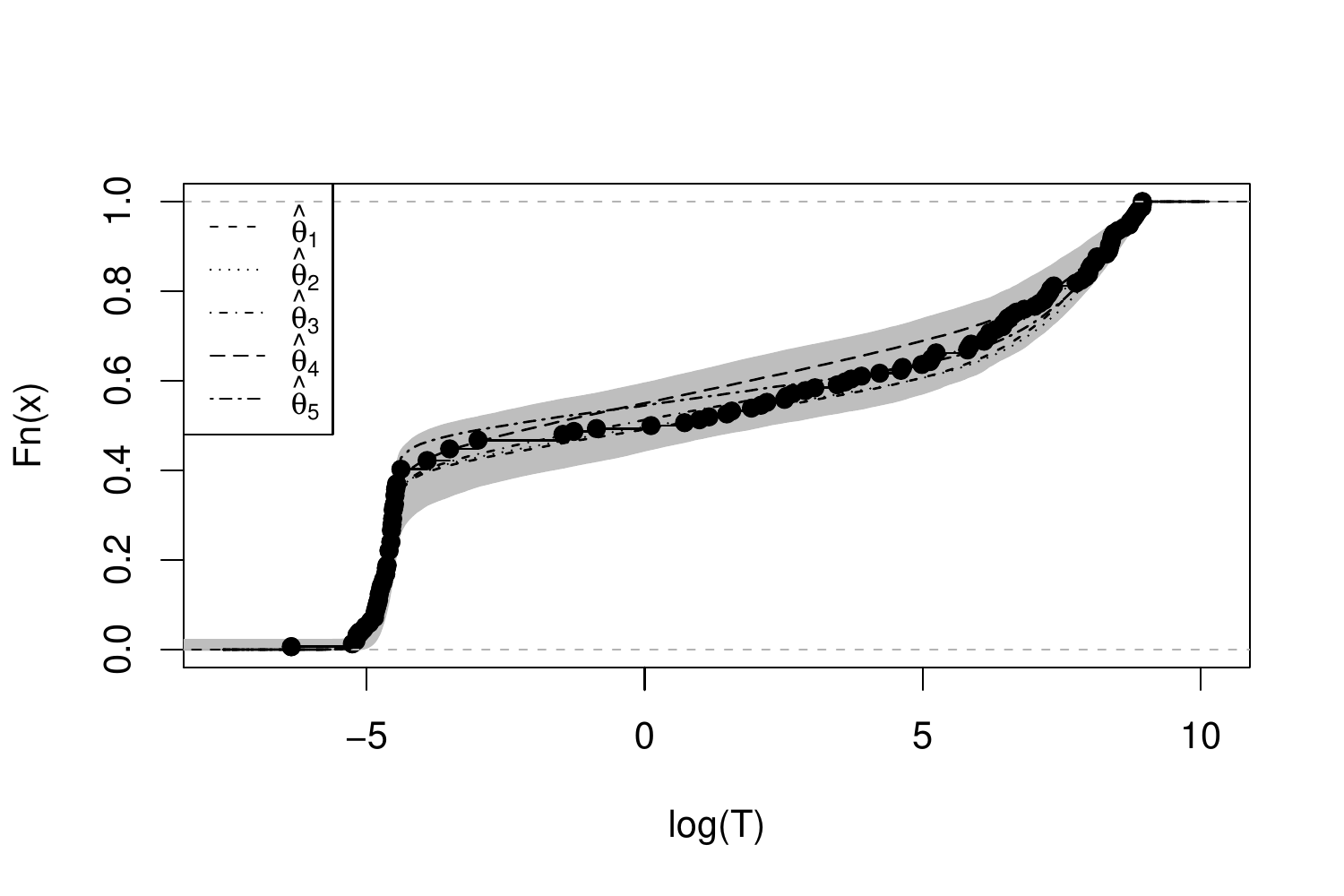}
\caption{Empirical CDF of the real data, and CDFs computed from the parameter vectors in Table \ref{table: estimation_data}. The top panel is the 3000/4Y dataset and the bottom panel is the 4500/1Y dataset. The gray area is the $95\%$ posterior interval of the estimated empirical CDFs.}
\label{fig:ecdf_data}
\end{figure}

To analyze the data, we followed the same procedure as described in Section \ref{sect:simulation} for multiple datasets and chose $\delta = 1.3$, which gave an overall ABC-MCMC acceptance rate of $0.88\%$.  To set starting values for the algorithm, we used the NLS estimates from \cite{foschiyao1986dol} as guidance, modified according to our parametrization. Table \ref{table: estimation_data} shows the five parameter vectors with the highest log-likelihood values. The histogram and the empirical cumulative distribution function (ecdf) of the data, along with the corresponding smoothed densities and CDFs for the parameter vectors from Table \ref{table: estimation_data} are shown in Figure \ref{fig:density_data} and Figure \ref{fig:ecdf_data}.  The results show that these parameter vectors indeed provide a very good fit of the data, and capture the variability in the individual parameters.

\begin{center}
\begin{table}[!htbp]
\caption{Five sampled parameter vectors, log-likelihoods, and log-posteriors for the real data.}
\begin{tabular}{ c|cccccccccc|c|c}
                 & $\mu_{a}$ & $\sigma_{a}$ & $\mu_{b}$ & $\sigma_{b}$ & $\mu_{c}$ & $\sigma_{c}$ & $\mu_{n}$ & $\sigma_{n}$ & $\mu_{\sigma_0}$ & $\sigma_{\sigma_0}$ & $ll$ & log-$post$ \\
\hline
$\hat{\theta}_1$ & -7.76 & 0.48 & 3.21 & 0.18 & -21.96 & 0.29 & -1.00 & 0.20 & 0.15 & 0.07 & -1120.37 & -1144.36 \\
$\hat{\theta}_2$ & -7.68 & 0.44 & 3.23 & 0.10 & -22.12 & 0.10 & -0.99 & 0.15 & 0.29 & 0.16 & -1122.37 & -1145.56 \\
$\hat{\theta}_3$ & -7.98 & 0.49 & 3.29 & 0.14 & -27.05 & 0.13 & -1.10 & 0.32 & -0.15 & 0.20 & -1123.51 & -1147.81 \\
$\hat{\theta}_4$ & -7.88 & 0.42 & 3.33 & 0.11 & -17.72 & 0.19 & -0.39 & 0.26 & -0.13 & 0.19 & -1123.76 & -1147.29 \\
$\hat{\theta}_5$ & -7.66 & 0.45 & 3.45 & 0.07 & -16.46 & 1.12 & -0.56 & 0.08 & 0.31 & 0.18 & -1123.81 & -1148.49
\end{tabular}
\label{table: estimation_data}
\end{table}
\end{center}

\section{Assessing long-term lumber reliability} \label{sect:prediction}

\subsection{Reliability analysis for live loads}

The key application of the fitted ADM is the assessment of the long-term reliability of a piece of lumber under stochastic loadings. The key element in this assessment is the performance equation
\begin{equation}\label{eq:performanceeq}
	G = C - D
\end{equation}
where both the future demand to be made on a random piece of lumber $D$ and its capacity to meet that demand $C$ depend on a random vector of random design variables that after a suitable transformation have a standard multivariate normal distribution. A Laplace approximation yields the probability of failure
$$p_f = P(G \leq 0) \approx 1 - \Phi(\beta)$$ where $\beta$ is called the \emph{reliability index}.  See \citet{madsen2006methods} for more details.

The random future load to which that piece will be exposed is the sum of two random components, the dead load $D_d$ and the live load $D_l$.  For illustrative purposes, we adopt the generative model of future loads for $\tau(t),~t \geq 0$ presented in \cite{foschi1989reliability}.   We briefly review the stochastic models with which the future loads are simulated and how those load levels are used to construct plots for reliability assessment.

At the basis of the model are certain specified constants called design values, which are in the National Building Code of Canada (NBCC) standards document CAN/CSA-O86: $d_{nd}$ and $d_{nl}$. The design values for the dead and live loads are then modelled as the constants $d_d = \alpha_d d_{nd}$ and $d_l = \alpha_l d_{nl}$, respectively for specified parameters $\alpha_d = 1.25$ and $\alpha_{l} = 1.5$. The design load is the constant $d_d + d_l$.

The corresponding constant for capacity $C$ is based on the characteristic value $R_o$ for a given lumber population, which in our example will be the fifth percentile of the strength distribution of a hemlock species $R_o = 2722$ psi. The design capacity is then $\phi^\prime R_o$ for some constant $\phi^\prime$, with corresponding design performance $ \phi^\prime R_o - (d_d + d_l)$. The design capacity will equal or exceed the design demand if
$\phi^\prime$ is set at the $ \phi$ for which
\begin{equation}\label{eq:designperformance}
\phi R_o - (d_d + d_l)= 0.
\end{equation}
Another design value of importance is the dead to live load ratio $\gamma = d_{nd}/d_{nl}$, which is typically $0.25$. A little algebra then shows
\begin{equation}\label{eq:liveload}
	d_{nl} = \frac{\phi R_o}{\gamma \alpha_d + \alpha_l}.
\end{equation}

The parameter
$\phi$ is called the \emph{performance factor} and like $\beta$ plays a fundamental role in reliability modelling. For a strong species it will be small while for a weak species it will be large. Clearly $\beta$ and $\phi$ must be related since a large value of the former would mean a small chance of failure, which in turn would mean a strong population and a small value of $\phi$.

However in reality the dead and live loads are random and their distributions must now be specified.  This is done by using the design values as a baseline and normalizing the loads as $\tilde{D}_d = D_d/d_{nd}$ and $\tilde{D}_l = D_l/d_{nl}$.  We will confine our analysis in this article to live residential loads, which is one of the many cases explored in \citet{foschi1989reliability}.  Hence we adopt their stochastic load specifications, by first assuming that
$\tilde{D}_d \sim N(1,0.01)$,
which is constant for the life of the structure.

The live loads are modelled as a sum of loads from two independent processes: sustained and extraordinary.
The sizes of the loads are modelled using gamma distributions  $G(k,\theta)$ where $k$ and $\theta$ represent the shape and scale parameters.  The random times between and during live load events are modeled using exponential distributions $Exp(\lambda)$ with mean $\lambda^{-1}$. Parameters for these models were previously fitted using survey data  \citep{corotis1977probability,chalk1980probability,harris1981area}. Hence the normalized live load at time $t$ is given by the stochastic process $\tilde{D}_l(t) = \tilde{D}_s(t) + \tilde{D}_e(t)$, where $\tilde{D}_s$ and $\tilde{D}_e$ are the normalized sustained and extraordinary loads respectively.

The process $\tilde{D}_s(t)$ consists of a sequence of successive periods  of sustained occupancy each with iid duration $T_s \sim Exp(1/0.1)$.  During these periods of occupancy $\tilde{D}_{ls} \sim G(3.122,0.0481)$ iid.  The process $\tilde{D}_e(t)$ consists of brief periods of extraordinary loads, separated by longer periods with no load $T_e\sim Exp(1.0)$ of expected duration one year.  When extraordinary loads occur, they last for iid periods of random duration  $T_p\sim Exp(1/0.03835)$. The normalized loads $\tilde{D}_{le}$ during these brief periods are iid with gamma distribution $\tilde{D}_{le} \sim G(0.826,0.1023)$.

\begin{figure}[!htbp]
\centering
\includegraphics[scale=0.8]{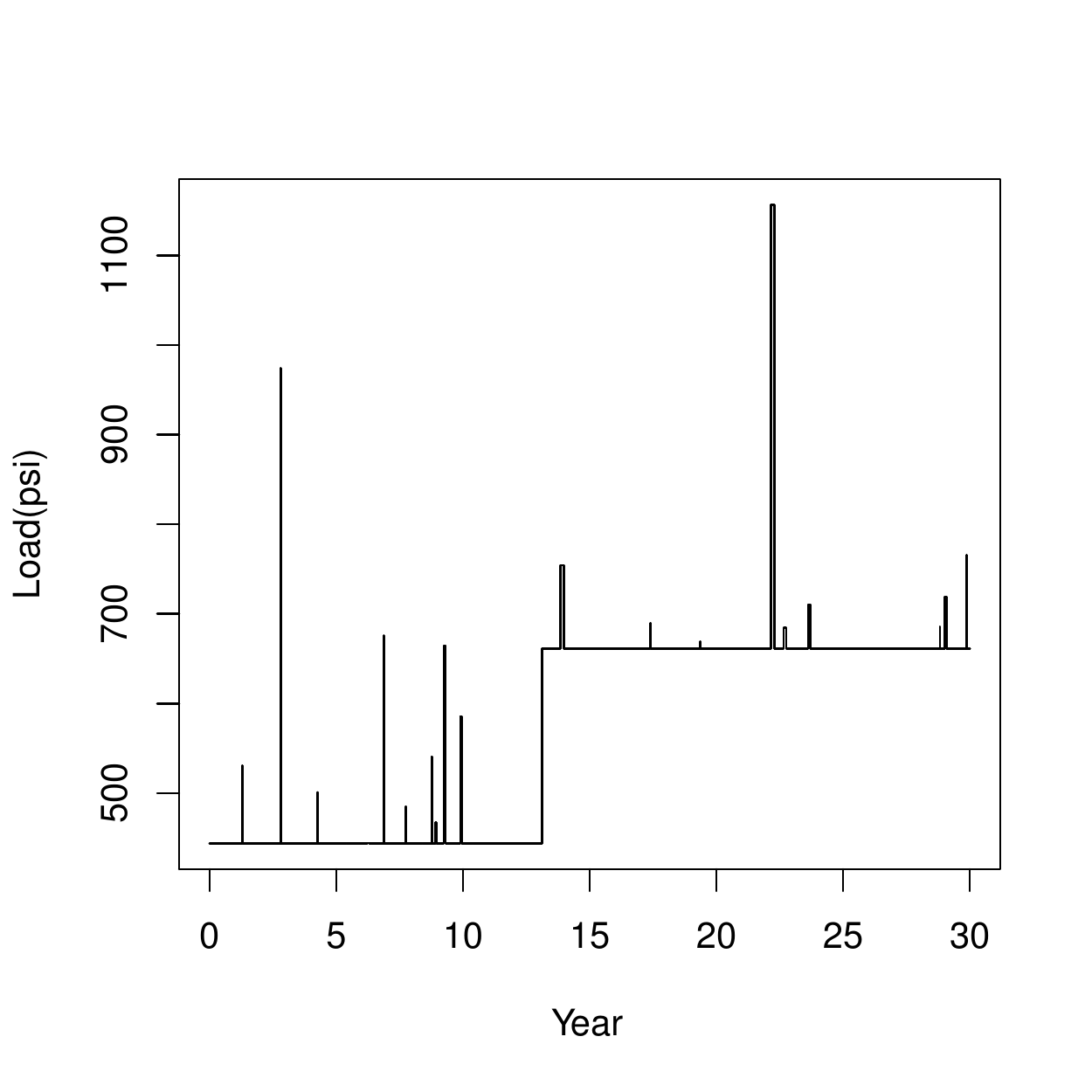}
\caption{An example residential load profile with $\phi = 1$.}
\label{fig:load}
\end{figure}

The combined normalized dead and live loads are then converted to actual load levels $\tau(t)$.  Applying Equation
(\ref{eq:liveload}) it is easily shown that
\begin{equation}\label{eq:deadpluslive}
\tau(t)=\phi R_o\frac{\gamma\tilde{D}_d + \tilde{D}_s(t) + \tilde{D}_e(t)}{\gamma\alpha_d  + \alpha_l}.
\end{equation}
An example of a simulated 30-year load profile according to these settings is shown in Figure \ref{fig:load}.

For a reliability analysis, the required service time for a piece of lumber is assumed to be $30$ years.  To simulate the damage accumulation process for a single random specimen, its random effects are sampled along with a realization of $\tau(t)$.  Then according to the ADM, the specimen is deemed to have failed if for some $t\leq 30$, $1\leq \alpha(t)$ and the associated survival time would be the smallest $t$ for which that is true.  Based on a large number of replications, the estimated probability of failure after $30$ years is obtained, $\hat{p}_f$.  From that estimate we compute
	\[
	\hat{\beta} = -\Phi^{-1}(\hat{p}_f)
	\]
where $\Phi$ is the standard Normal CDF.  By repeating this procedure for successive values of $\phi$, the functional relationship between $\beta$ and $\phi$ can be estimated.

\subsection{Analysis Results}

The Canadian model cannot be solved analytically for an arbitrary loading profile $\tau(t)$, so we obtain its solution using the \texttt{odeint} in C++ Boost library. This library provides a wide range of ODE solvers and we use the five-step Adams-Bashforth method for the sake of efficiency.

Based on our Bayesian framework, the posterior distribution of the future time-to-failure $T_f$ can be estimated using the MCMC samples $\theta_i$ of $\theta$,
\begin{align*}
p(t_f | \boldsymbol{t}_{obs}) &\approx \sum_{i=1}^{n_i} \sum_{j=1}^{n_j} p( t_f |  \theta_i, \tau_{ij}(t))\\
&\approx \sum_{i=1}^{n_i} \sum_{j=1}^{n_j} p( t_f |  a_{ij}, b_{ij}, c_{ij}, n_{ij}, \sigma_{0,ij}, \tau_{ij}(t))
\end{align*}
where each $\tau_{ij}(t)$ is an independent realization of the stochastic load profile, and \\
$a_{ij}, b_{ij}, c_{ij}, n_{ij}, \sigma_{0,ij}$ are independent draws of the piece-specific random effects (\ref{eqn:random-effect}) conditioning on $\theta_i$.

\begin{figure}[!htbp]
\centering
\includegraphics[scale=0.8]{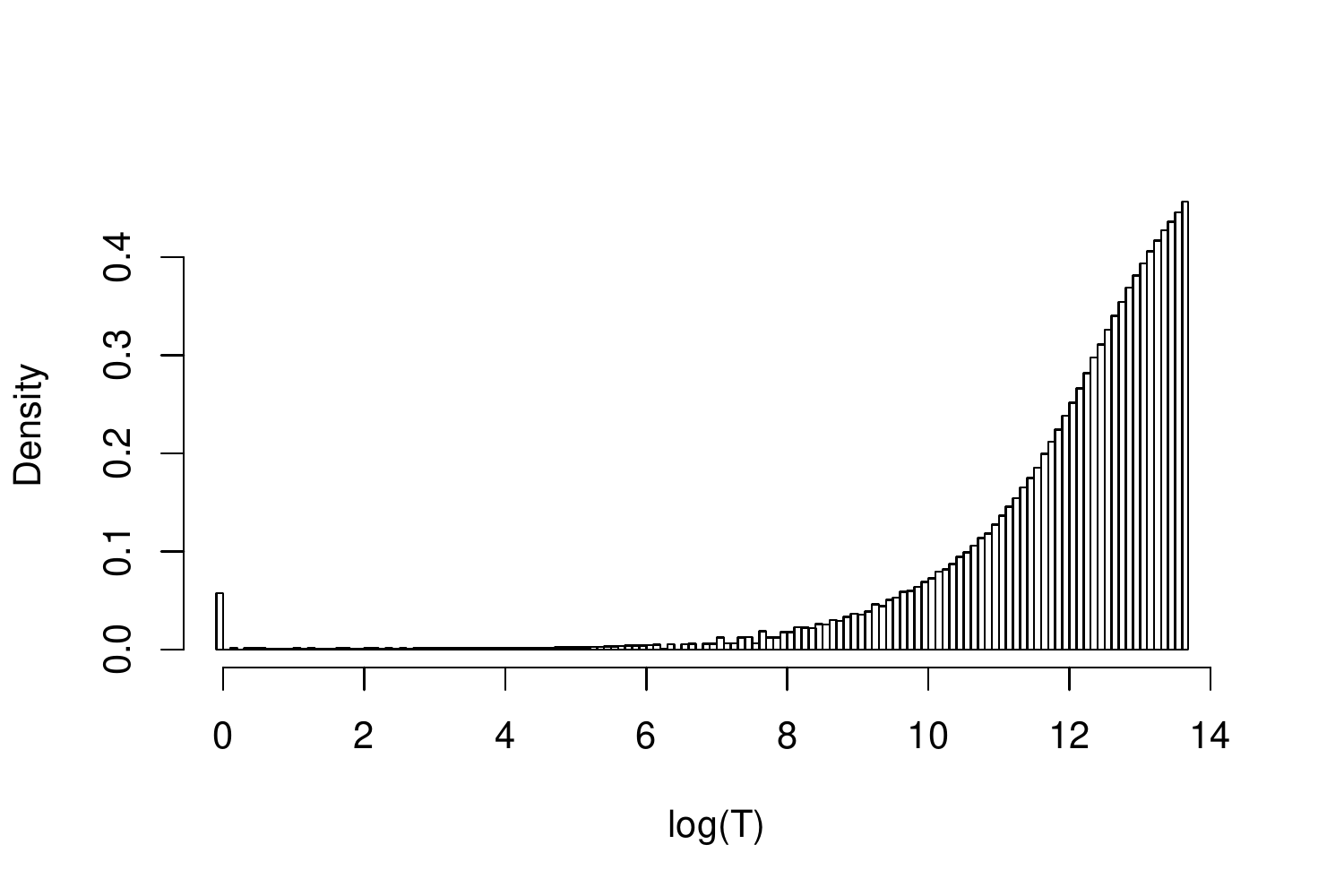}
\caption{Histogram of the posterior distribution of  time-to-failure with a $\phi = 3$ load profile, for failures that occur within the first 100 years.}
\label{fig:hist_100y}
\end{figure}

Hence our simulation procedure is as follows: for each of the $n_i=500$ draws of $\theta$ in Section \ref{sect:analysis}, we generate $a, b, c, n, \sigma_0$ using (\ref{eqn:random-effect}). Then we solve the Canadian ADM for the time-to-failure $T_f$ with this $a, b, c, n, \sigma_0$ and a randomly generated load profile (\ref{eq:deadpluslive}) with the given $\phi$. We replicate this $n_j=100,000$ times for each draw of $\theta$.  For example, the posterior distribution of the time-to-failure $T_f$ given the data $\boldsymbol{t}_{obs}$ with a $\phi = 3$ load profile is shown in Figure \ref{fig:hist_100y}. Note that there is a small peak at the bottom of the histogram; these correspond to the weakest pieces of lumber that do not survive the initial loads under this scenario.

\begin{figure}[!htbp]
\centering
\includegraphics[scale=0.6]{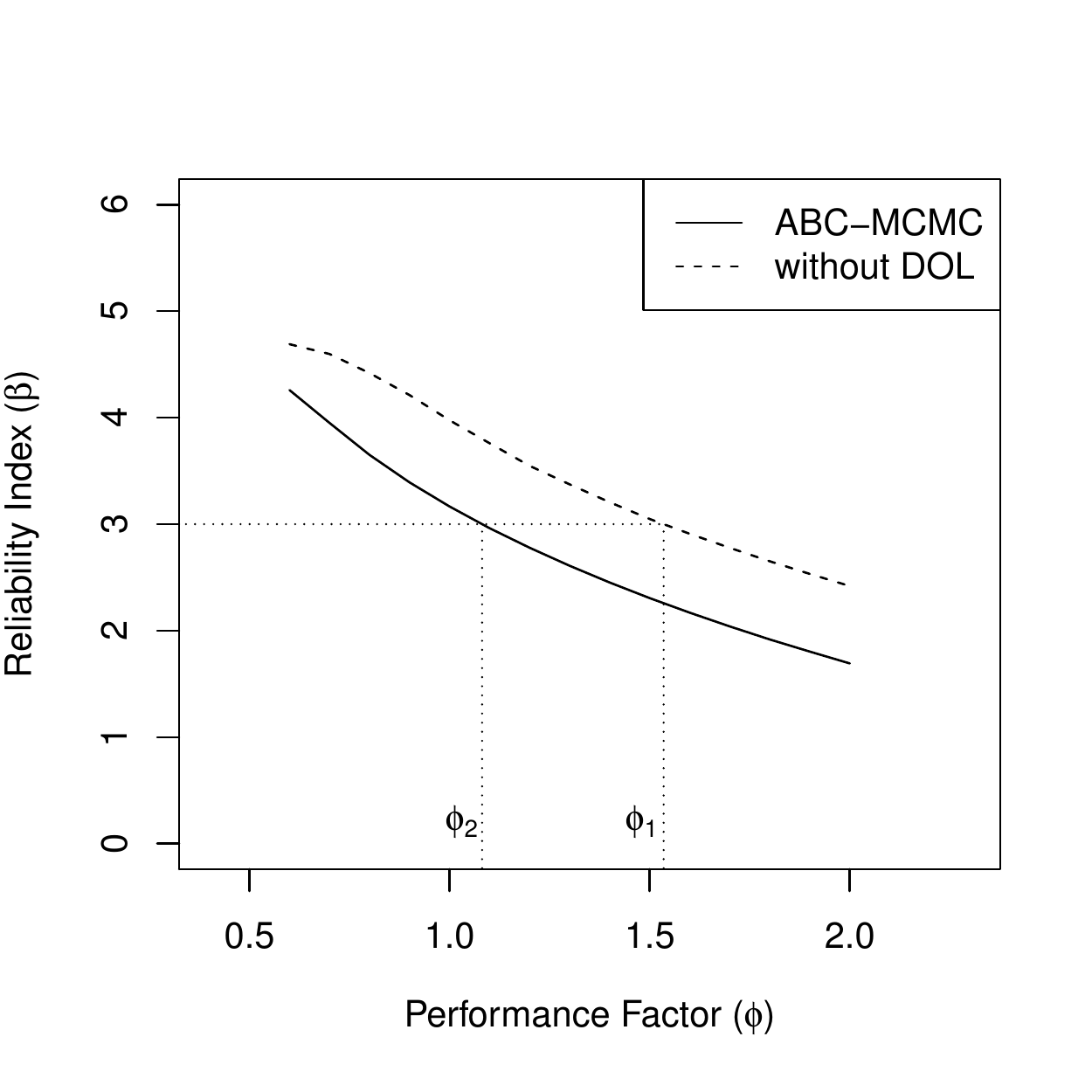}
\includegraphics[scale=0.6]{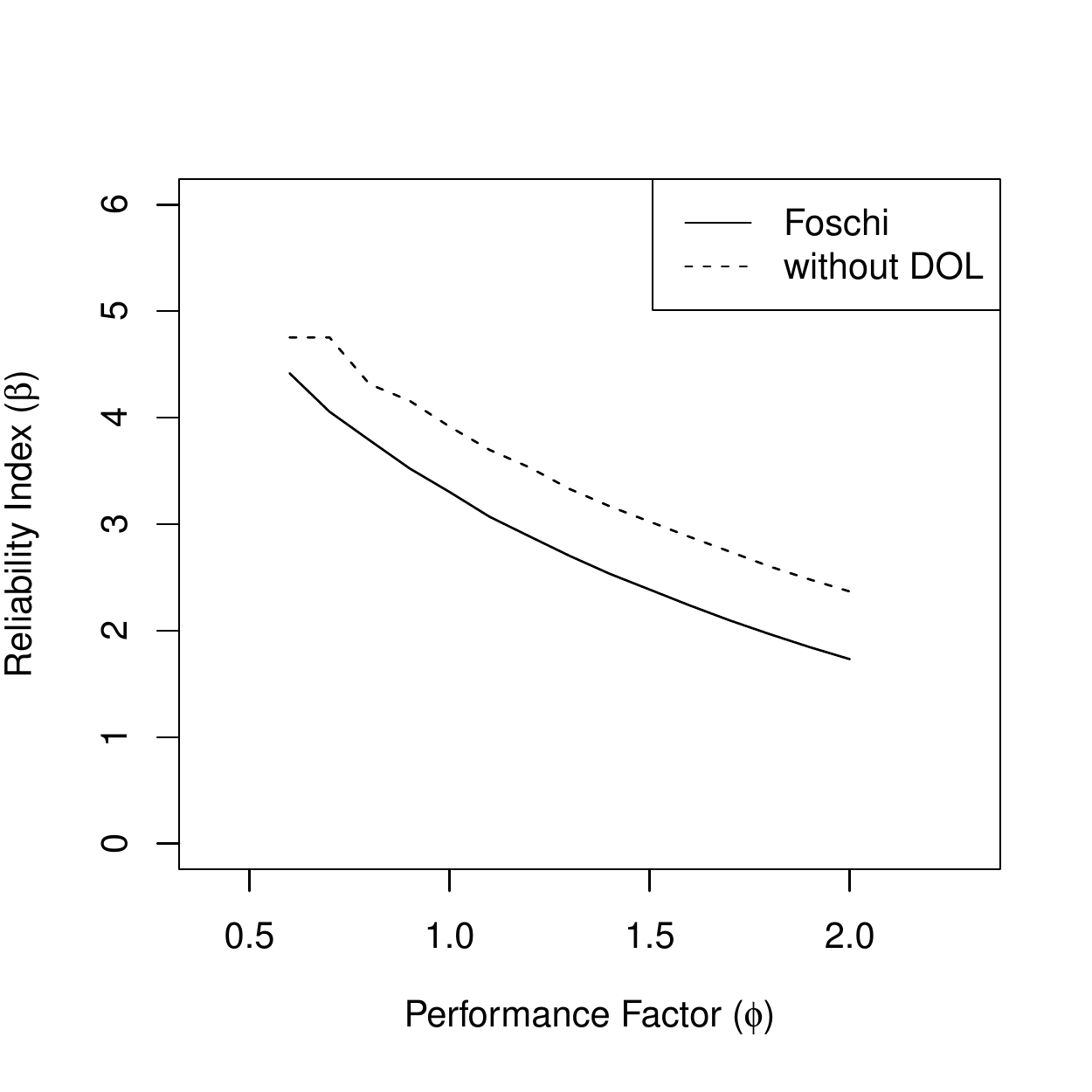}
\caption{$\phi - \beta$ relationship for ABC-MCMC and Foschi's estimates.}
\label{fig:reliability}
\end{figure}

This procedure provides the estimated probability of failure $\hat{p}_f$ by the end of 30 years and the associated reliability index $\hat{\beta}$. To quantify the DOL effect, we also calculate the reliability index assuming there is no DOL effect. When there is no DOL effect, a piece of lumber breaks if the maximum load exceeds its short-term strength $\tau_s$ during the 30-year period. The result is shown in Figure \ref{fig:reliability}. We also replicate the result in \cite{foschi1989reliability} by generating $a, b, c, n, \sigma_0$ using their estimates and parametrization.  Then for a fixed $\beta$, we can measure the DOL effect by taking the ratio of the two corresponding performance factors $\phi_1$ and $\phi_2$ as indicated for $\beta=3$ in Figure \ref{fig:reliability}. \cite{foschi1989reliability} define this ratio as the adjustment factor $K_D$, i.e.
\[
K_D = \frac{\phi_2}{\phi_1},
\]
where $\phi_1$ and $\phi_2$ are the performance factors corresponding to the specified value of $\beta$ when DOL effect is absent and present, respectively. The result is shown in Table \ref{table: phi-ratio}. For our method, we are also able to calculate the $95\%$ posterior interval for $K_D$ using the MCMC samples.  The point estimates shown for our approach are the posterior means.

\begin{table}[!htbp]
\begin{center}
\caption{The adjustment factors $K_D$ for ABC-MCMC and Foschi's estimates.}
\begin{tabular}{ c|cccc|ccc}
              & \multicolumn{4}{c|}{ABC-MCMC}                        & \multicolumn{3}{c}{Foschi}\\	
\hline
              & $\phi_{2}$ & $\phi_{1}$ &  $K_D$ &  $95\%$ Interval & $\phi_{2}$ & $\phi_{1}$ &  $K_D$ \\
\hline
$\beta = 2.5$ & 1.37 & 1.93 & 0.71 & (0.56, 0.81) & 1.42 & 1.88 & 0.76 \\
$\beta = 3.0$ & 1.08 & 1.53 & 0.71 & (0.53, 0.81) & 1.14 & 1.52 & 0.75 \\
$\beta = 3.5$ & 0.86 & 1.23 & 0.70 & (0.49, 0.82) & 0.91 & 1.22 & 0.75 \\
\end{tabular}
\label{table: phi-ratio}
\end{center}
\end{table}

We find that our approach provides a more conservative estimate of reliability, while the result of \cite{foschi1989reliability} is well within the range that would be expected due to uncertainty from parameter estimation.

\section{Discussion and conclusions} \label{sect:discussion}

In this article, we presented a Bayesian framework for estimating the parameters and quantifying the uncertainty in the parameters for the Canadian ADM. We adapted an ABC algorithm to handle the computational challenges.  Using the fitted model, we presented an application to reliability analysis using the posterior distributions of the parameters.

Our approach provides posterior probability intervals that quantify the DOL effect, in particular the important adjustment factor $K_D$; such interval estimates could not be obtained by the approach in \citet{foschi1989reliability}.  Future work can extend this Bayesian framework to other forms of ADMs and exploring alternative or reduced parametrizations that can still fit the data well, since some of the parameters are highly uncertain in the posterior.  Other forms of future live loads can also be added to the analysis -- such as snow, wind, and earthquakes -- so as to obtain more realistic different stochastic load patterns for $\tau(t)$.

Our work also shows that ABC-MCMC indeed is a promising approach for complicated models. In this case, for a subset of sampled parameter vectors it is possible to directly assess and verify the goodness of the ABC approximation with a brute-force likelihood computation.  Hence, our adapted version of ABC-MCMC for censored data can be seen here as a useful computational device that helps to efficiently explore the parameter space and sample good candidates of the parameter vector.

\subsection*{Acknowledgements}
The work reported in this manuscript was partially supported by FPInnovations and a CRD grant from the Natural Sciences and Engineering Research Council of Canada.  The data analysed in this paper were provided by FPInnovations. We are greatly indebted to Conroy Lum and Erol Karacabeyli from FPInnovations for their extensive advice during the conduct of the research reported herein.

\appendix

\section{Appendix: Canadian model derivation}
For a ramp load test using the standard loading rate $k_s$, then we have $\tau(t) = k_s t$, $\tau_s = kT_s$, and
  \begin{eqnarray*}\label{eq:canramp}
 	\frac{d}{dt}\alpha(t) \mu &=& [{a} k T_s(t/T_s - \sigma_0)_+]^b
  + [{c} k T_s (t/T_s - \sigma_0)_+]^n \alpha(t).
  \end{eqnarray*}
  Define the integrating factor
  \begin{eqnarray*}
  H(t) &=& \exp \left\{ \int -\frac{1}{\mu} \left[ {c} k T_s \left( \frac{t}{T_s} - \sigma_0 \right) \right] ^n \,dt \right\} \\
  &=& \exp \left\{ -\frac{1}{\mu}  ({c} k T_s)^n \frac{T_s}{n+1}  \left( \frac{t}{T_s} - \sigma_0 \right)^{n+1} \right\}.
  \end{eqnarray*}
  Then
  \begin{eqnarray*}
\frac{d}{dt} \left[ \alpha(t) H(t) \right]  = \frac{1}{\mu} \cdot H(t) \left[ {a} k T_s  \left( \frac{t}{T_s} - \sigma_0 \right) \right]^b.
  \end{eqnarray*}
No damage is accumulated until $t = \sigma_0 T_s$, so integrating we obtain
  \begin{eqnarray*}
  \alpha(T_s)H(T_s) - \alpha( \sigma_0 T_s)H(\sigma_0 T_s) = \int_{\sigma_0 T_s}^{T_s} \frac{1}{\mu} \cdot H(t) \left[ {a} k T_s  \left( \frac{t}{T_s} - \sigma_0 \right) \right]^b \,dt.
  \end{eqnarray*}
Finally the change of variables $u = -\log H(t)$ yields Equation (\ref{eq:Tssoln}), where we then recognize the integral to be the lower incomplete Gamma function, which can be evaluated numerically using standard mathematical libraries.

\bibliographystyle{asa}

\end{document}